\documentclass{PoS}
\usepackage{wrapfig}
\usepackage{epsfig}

\title{The ATLAS Forward Physics Program}

\ShortTitle{The ATLAS Forward Physics Program}

\author{\speaker{Christophe Royon}\\
        CEA/IRFU/Service de physique des particules, CEA/Saclay, 91191 
Gif-sur-Yvette cedex, France\\
        E-mail: \email{christophe.royon@cea.fr} \\
	On behalf of the ATLAS experiment}

\abstract{We describe the ATLAS Forward Physics Program at low luminosity using
the rapidity gap method and a dedicated detector called ALFA to tag the protons.
We also describe the physics topics of the ATLAS Forward Physics Project at high
instantaneous luminosity.}

\FullConference{XVIII International Workshop on Deep-Inelastic Scattering and Related Subjects, DIS 2010\\
		April 19-23, 2010\\
		Firenze, Italy}

\begin{document}

\section{ATLAS Forward Physics Program}
The ATLAS Forward Physics Program benefits from the good coverage of the ATLAS
detector in the forward region~\cite{brandt}. In particular, we can quote the
Lunminosity Cerenkov Integrating Detector (LUCID) at 17 m from the ATLAS nominal
interaction point, the Zero Degree Calorimeter (ZDC) at 140 m and the Absolute
Luminosity for ATLAS (ALFA) roman pots at 240 m. In addition, the ATLAS Forward
Physics Project (AFP) under discussion within ATLAS foresees to install
additional forward detectors (movable beam pipes) at 220 and 420 m from the
ATLAS interaction point.

The first diffractive measurements which can be performed in ATLAS are given
in Fig.~1. At low luminosity, it is possible to select diffractive events using
the forward rapidity gap method. Since there is no colour exchange between the
intact proton in the final state and the object produced in the central region
(pions, jets, photon...), a rapidity gap devoid of any activity is present in
the forward region. The ATLAS forward detectors and their good coverage in the
forward region (the Forward Calorimeter FCAL
3.2$<|\eta|<$4.9, LUCID 5.6$<|\eta|<$6.0 and ZDC $|\eta|>$8.3) allow to measure 
single diffraction and double pomeron exchanges. The central gaps can be
measured using the hadronic calorimeter $|\eta|<$3.2) and 
the inner detectors ($|\eta|<$2.5). More complicated events such as the last one
of Fig.~1 can also be measured.

In the next sections, we will cover the potential measurements from ATLAS using
the rapidity gap method, the ALFA roman pots at low luminosity and the AFP
movable beam pipes at high luminosity.

\begin{figure}
\begin{center}
\includegraphics[width=0.8\textwidth]{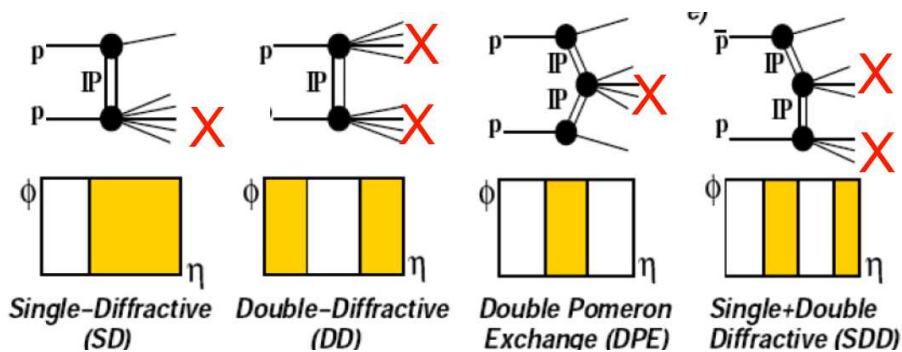}
\caption{\label{schemedif}
Scheme of diffractive events.}
\end{center}
\end{figure}

\section{Early diffractive measurements in ATLAS}
\subsection{Hard single diffraction and double pomeron exchanges}
One of the first possible measurements is to look for single diffractive events
where jets, $W$s or $Z$s are produced in the central detector using the rapidity
gap method. The soft survival probability can be determined using the first data
while comparing the data with or without a gap in the forward region. The first
diffractive measurements can be performed with a limited luminosity. As an
example, approximately 5000 (8000) single diffractive dijet events can be 
produced in 100 pb$^{-1}$ with a jet transverse momentum above 20 (40) GeV after
taking into account the trigger prescale at low luminosity.

Requesting two central jets in the central ATLAS detector and the presence of
a rapidity gap in each proton direction will allow to select Double Pomeron
Exchange events. As performed by the CDF collaboration at the
Tevatron~\cite{cdf}, it is possible to measure the dijet mass fraction as an
example, which allows to distinguish between exclusive and inclusive diffractive
events. The dijet mass fraction is defined as the ratio of the dijet mass and
the total mass in the event (measured for instance in the ATLAS calorimeter).
For exclusive events, the dijet mass fraction is close to 1 since only two jets
and the two scattered protons and nothing else are produced in this kind of
events. For inclusive events, part of the energy is lost in pomeron remnants and
the dijet mass ratio will be significantly smaller than 1. A such measurement
allows to measure the exclusive diffractive dijet production cross section and
it will be useful to constrain further the exclusive models and give better
predicton on diffractive exclusive Higgs boson production cross section in
particular~\cite{rafal}.

\subsection{Photon induced processes}
Two kinds of photon induced processes are specially interesting, namely the
exclusive dilepton production and the photoproduction processes. In exclusive
dilepton production $pp \rightarrow pllp$, two protons and two leptons 
originating from QED processes are
produced in the final state. To select such
events, on can require the presence of one rapidity gap on each proton side, two
isolated back-to-back leptons, the presence of an exclusive vertex (no other
track is present than those originating from the leptons). The typical cross
section is 10 pb for lepton $p_T$ above 10 GeV.

The photoproduction processes can produce $J/\Psi$ or $\Upsilon$ resonances
originating from photon-pomeron exchanges. The cross section is also of the
order of 10 pb and the processes can be detected via the leptonis decays of
$J/\Psi$ or $\Upsilon$. These events are particularly interesting to constrain
further the unintegrated gluon distributions which are one of the inputs to
compute diffractive exclusive cross sections, for instance for Higgs
production~\cite{martin}.

\subsection{Jet gap jet events}
The other process of interest which can be studied using the first ATLAS data is
the jet gap jet event. To select such processes, one requires the presence of
two jets reconstructed in the ATLAS calorimeter and a rapidity gap devoid of any
activity between them. These processes allow a direct test of the Balitsky Fadin
Kuraev Lipatov~\cite{bfkl} (BFKL) resummation and recently, the Next-to-Leading
Logarithm (NLL) BFKL equation  was implemented for these processes in
HERWIG~\cite{cyrille}. It leads to a fair description of the CDF and D0 data.

\section{Diffractive measurements using ALFA}
The main motivation of installing the ALFA detectors is the total cross section
measurement. This detector was described in another contribution at this
conference~\cite{brandt}. The idea is to measure the elastic cross section in
the Coulomb and interference region (see Fig.~3), which can be used to have an
absolute measurement of the luminosity. The elastic cross section is
the sum of the coulombian, nuclear and interference terms
 \begin{eqnarray}
\frac{dN}{dt} = L \left( \frac{4 \pi \alpha^2 G^4(t)}{|t|^2} -
\frac{\alpha \rho \sigma_{tot} G^2(t) e^{- B |t|/2}}{|t|} 
+ \frac{\sigma^2_{tot} (1+ \rho)^2 e^{- B |t|}}
{16 \pi} \right) .
\end{eqnarray}
The luminosity $L$, the total cross ection, and the $B$ and $\rho$ parameters
appearing in the elastic cross section formula are determined by fitting the
$dN/dt$ spectrum in the interference and nuclear regions~\cite{alfa}.
The measurement requires the possibility to detect the protons in the final
state down to $t\sim
3.7~10^{-4}$ GeV$^2$ which means a proton angle down to 3 $\mu$rad, which
requires special high $\beta^*$ runs at low luminosity. The total uncertainties
on the elastic cross section measurement are expected to be less than 3\% (
beam properties: 1.2\%, detector properties: 1.4\%, background substraction: 
1.1\%,
1.8\% statistical error for 100 hours of measurement at low luminosity).

\begin{figure}
\begin{center}
\includegraphics[width=0.8\textwidth]{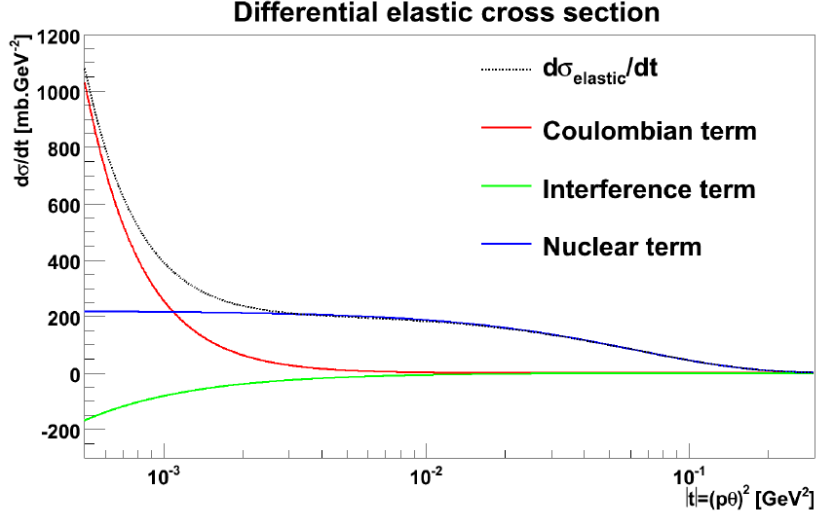}
\caption{\label{schemedifb}
Coulombian, nuclear and interference terms in the elastic cross section.}
\end{center}
\end{figure}

The ALFA detector also allows to measure soft single diffractive events in
dedicated runs where ALFA will be used to measure elastic events. It is possible
to measure forward protons in the region: $6.3<E_{proton}<7$
TeV, and single diffarctive measurements are possible 
for $\xi<0.01$ and non-diffractive proton
measurements for $0.01<\xi<0.1$.
1.5 million events are expected in 100 hours at 10$^{27}$
cm$^{-2}$s$^{-1}$.

\section{Diffractive measurements at high luminosity}
The AFP project under discussion in the ATLAS collaboration will allow to detect
protons in the final state using additional proton taggers to be intalled at 220 and
420 meters from the ATLAS nominal interaction point. The movable beam pipes will
host 3D Silicon detectors allowing to measure the position of the scattered
protons with a precision better than 10 $\mu m$ and time of flight detectors
(GASTOF and QUARTIC) to measure the arrival time of the protons with a precision
of 5-10 ps~\cite{brandt}.

In addition to QCD studies of diffractive events and a better understanding of
the pomeron structure, the main motivations of AFP are the exclusive diffractive
Higgs production and the study of $\gamma W$ and $\gamma Z$ anomalous couplings.
The exclusive Higgs production cross section and signal-over-background at the
LHC was studied in great details~\cite{higgs} after a full simulation of signal
and background events in the
ATLAS detectors. As an example, the sugnificance is larger than 3.5$\sigma$
(resp. 5$\sigma$) for 60 fb$^{-1}$ (resp. three years at the highest
luminosity). Diffractive Higgs production is complementary to the standard
non-diffractive search and allows a spin determination of the Higgs boson.

The study of $W$ and $Z$ pair production via photon exchanges ($pp \rightarrow
pWWp$) allows to study in detail the quartic and triple gauge anomalous $W\gamma$ and $Z \gamma$
couplings which are predicted in particular by Higgsless and extradimension
models. The present LEP limits on quartic anomalous couplings 
can be improved by up to four orders of magnitude by tagging the intact protons
in the final state and the $W$ and $Z$ decays into leptons for instance in the
ATLAS detector, which allows to reach the expected anomalous couplings for Higgsless
models~\cite{kepka}. The tagging of the protons using the ATLAS Forward
Physics detectors is the only method at present to test so small values of
quartic anomalous couplings and thus to probe the Higgsless models in a clean
way. In addition, photon-exchange processes allow to probe 
SUSY particle production and to asses their kinematical
properties~\cite{kepka}.
 
\begin{figure}[t]
\hfill
\begin{minipage}[t]{.45\textwidth}

\epsfig{file=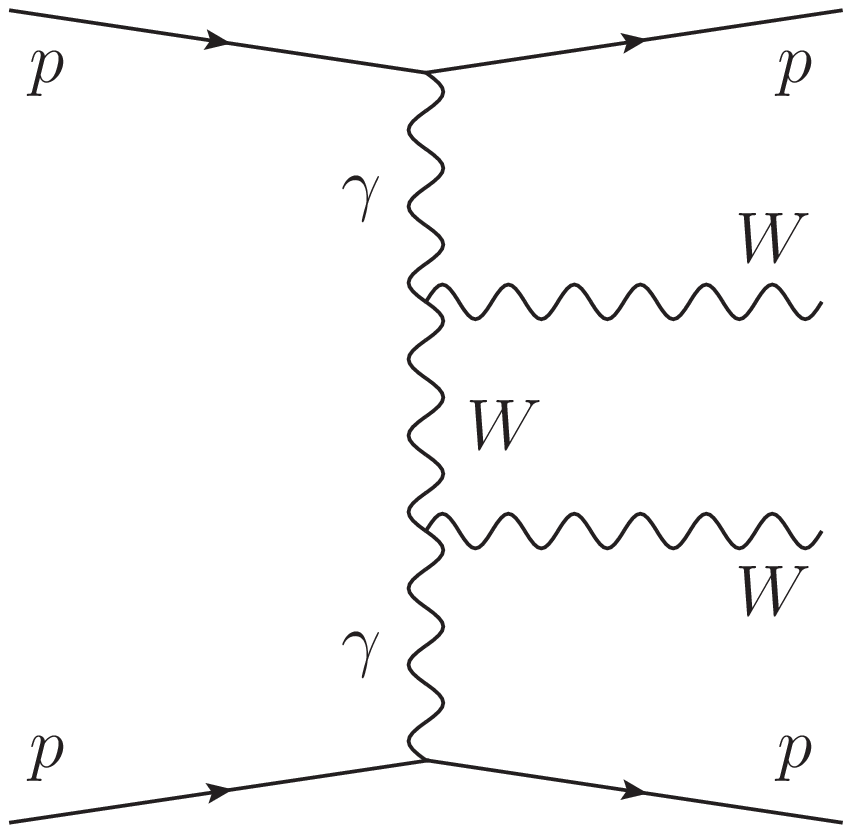,width=6.cm} 
\caption{Sketch diagram showing the two-photon production of a central system.}
\label{F2_QF_1}

\end{minipage}
\hfill
\begin{minipage}[t]{.45\textwidth}

\epsfig{file=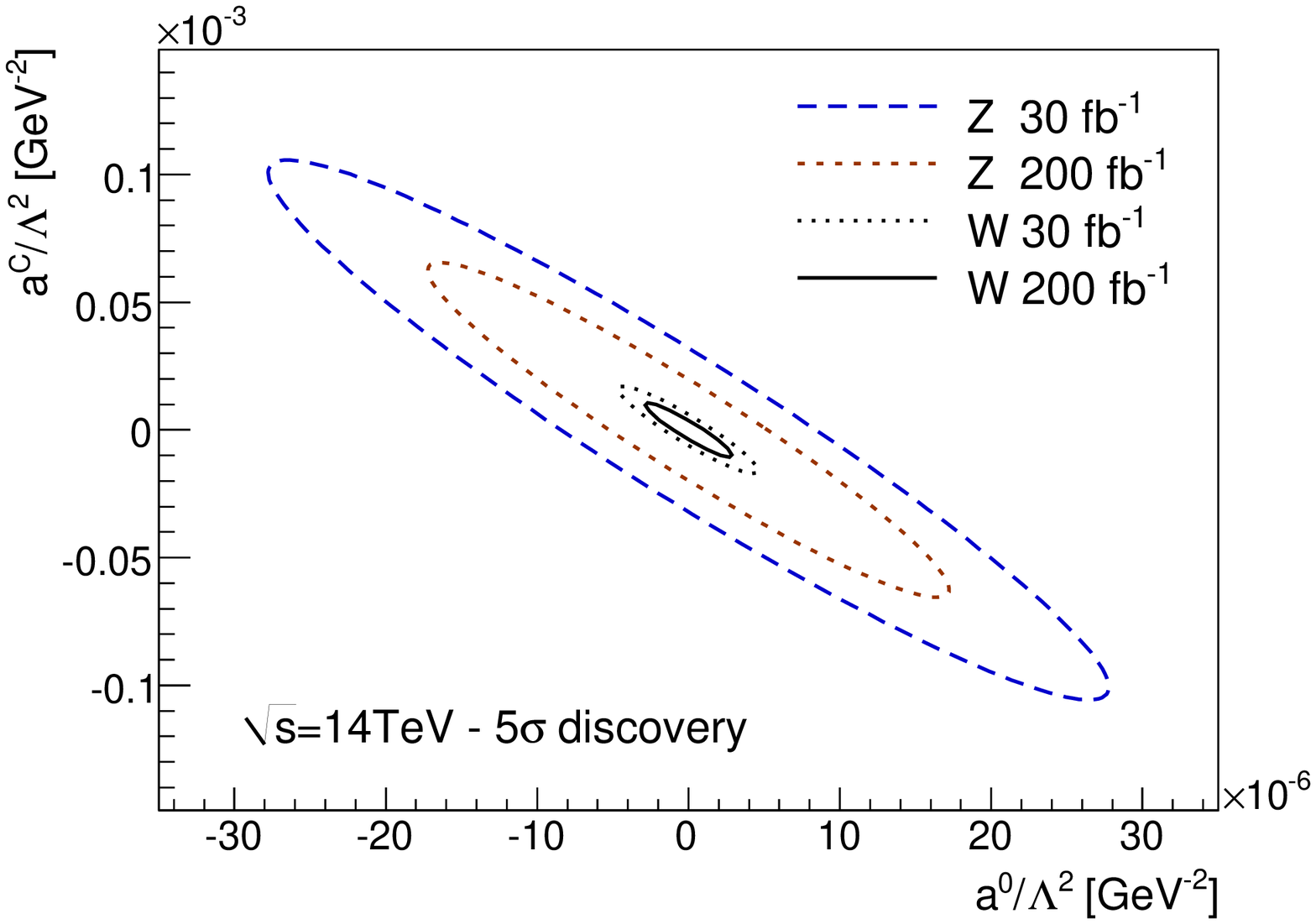,width=6.cm} 
\caption{$5\sigma$ discovery contours for all the $WW$ and $ZZ$ quartic 
couplings at $\sqrt{s}=14$ TeV for a luminosity of 30 fb$^{-1}$ and 200
fb$^{-1}$.}
\label{DVCS}

\end{minipage}
\hfill
\end{figure}

\section{Conclusion}
As a conclusion, we give in Table 1 the list of diffractive processes which can
be measured in ATLAS as a function of luminosity. The future of diffractive
measurements at the LHC is particularly rich and will especially benefit from
the AFP project which will be at the interface of standard QCD measurements,
beyond standard model searches and the search for the Higgs boson.

\begin{table} 
\begin{center}
\begin{tabular}{|c||c|} \hline
Luminosity& Possible measurements \\
\hline\hline
10 pb$^{-1}$ & Jet gap jet (Mueller Navelet) \\
  & Soft single diffraction \\
   & total cross section (ALFA) \\
   & Hard Single diffraction (jets, b jets...) \\ \hline
 10-100 pb$^{-1}$  & Central exclusive production (jets) \\
    & Single diffractive $W/Z$ \\ \hline
    100-200 pb$^{-1}$ & $WW$ via photon exchange \\
     & dilepton production \\
     & CEP $\tau\tau$ \\ \hline
 30 fb$^{-1}$ & Higgs (with AFP) \\
  & Anomalous $W\gamma$ couplings  (with AFP) \\
  & Test of Higgsless / extradim models (with AFP)\\
\hline
\end{tabular}
\caption{Possible diffractive measurements in ATLAS as a function of accumulated
luminosity}
\end{center}
\end{table}

\end{document}